\documentstyle[12pt,epsf]{article}
\begin{document}
%
\newcommand{\nc}{\newcommand}
\def\veffT{\Delta V_{{\rm eff},T}}
\def\msh{m^2_{h^0}}
\def\veffo{V_{{\rm eff},0}}
\nc{\be}[1]{\begin{equation} \mbox{$\label{#1}$}}
\nc{\ee}{\end{equation}}
\nc{\bea}[1]{\begin{eqnarray} \mbox{$\label{#1}$}}
\nc{\eea}{\end{eqnarray}}
\nc{\lra}{\leftrightarrow}
\nc{\sss}{\scriptscriptstyle}
{\nc{\lsim}{\mbox{\raisebox{-.6ex}{~$\stackrel{<}{\sim}$~}}}
{\nc{\gsim}{\mbox{\raisebox{-.6ex}{~$\stackrel{>}{\sim}$~}}}
\def\dsl{\partial\!\!\!/}
\def\lameff{\lambda_{\rm eff}}
\def\Re{{\rm Re\,}}
\def\Im{{\rm Im\,}}
\def\wrt{{{\em w.r.t.\ }}}
\def\diag{{\rm diag}}
\def\sign#1{{\rm sign}(#1)}
\def\VEV#1{{\langle #1 \rangle}}
\def\nBR#1{{\bigl( #1 \bigr)}}
\def\NBR#1{{\left( #1 \right)}}
\def\DBR#1#2{\bigl\{#1 \bigr\}\bigl\{#2 \bigr\}} 
\def\VBBR#1{{\Biggl[ #1 \Biggr]}}
%
%
\begin{titlepage}
\pagestyle{empty}
\baselineskip=21pt
\rightline{NORDITA-99/67 HE}
\rightline{NBI-HE-99-42, UNIL-IPT/99-3}
\rightline{LPT Orsay 99-86}
\rightline{October 1999}
\vskip .5in
\begin{center}
      {\Large {\bf Quantum Transport Equations for a Scalar Field}}
\end{center}
\vskip .3in
\begin{center}
      Michael Joyce \\
        {\it LPT, Universit\'e Paris-XI, B\^atiment 211,
F-91405 Orsay Cedex, France, and  \\
             INFN, Sezione di Roma 1, Italy} \\
      Kimmo Kainulainen\\
        {\it NORDITA,
             Blegdamsvej 17, DK-2100, Copenhagen \O , Denmark}\\
      Tomislav Prokopec\\
      {\it Universit\'e de Lausanne, Institut de physique th\'eorique,
BSP, CH-1015 Lausanne, Suisse}\\
\end{center}
\vskip 0.5in
\centerline{\bf Abstract}
\vskip 0.3truecm
\baselineskip=18pt
We derive quantum Boltzmann equations from Schwinger-Dyson equations
in gradient expansion for a weakly coupled scalar field theory with
a spatially varying mass. We find that at higher order in gradients
a full description of the system requires specifying not only an on
shell distribution function but also a finite number of its derivatives,
or equivalently its higher moments. These derivatives describe quantum
coherence arising as a consequence of localization in position space.
We then show that in the limit of frequent scatterings coherent quantum
effects are suppressed, and the transport equations  reduce
to the single Boltzmann equation for particle density, in which particles
flow along modified semiclassical trajectories in phase space.

\end{titlepage}
\newpage
%
%
%
\section{Introduction}
\label{sec: Introduction}
\baselineskip=20pt
In this letter we present a controlled derivation of dynamical transport
equations for a simple complex scalar theory
\be{Lagrange}
      {\cal L} =
       \NBR{\partial_\mu \phi}^{\dagger} \NBR{\partial^\mu \phi}
       - m^2 (\vec{x}, t) \phi^\dagger \phi + {\cal L}_{\rm int}\,,
\ee
where the mass represents coupling to a classical background field
which varies in space and time, and ${\cal L}_{\rm int}$ denotes
interactions. For definiteness we consider here the simple quartic
interaction ${\cal L}_{\rm int}=-\lambda(\phi^*\phi)^2/4$.
In particular we study the effect of the breakdown of translational
invariance with a treatment of the varying background mass at
nontrivial order in gradients.

The motivation for this work comes from electroweak baryogenesis
at a first order phase transition \cite{SR}, where one needs to model
the departure from thermal equilibrium induced at the phase boundary of
a growing bubble of the broken phase. No systematic treatment of
such plasma dynamics, adequate to incorporate the crucial CP violating
effects, is yet available. This paper is the second in a series
in which we attempt to provide a systematic formalism for the treatment
of this problem. Including higher order gradients is required since, in
realistic cases, the relevant CP violating effects occur typically
at higher order in gradients. The formalism we are developing is
general and may be applied to other problems which involve
analogous physics.

Our treatment of the problem is based on the out-of-equilibrium
closed time contour (CTC) formalism. In our derivation we assume a weak
coupling limit and the semiclassical approximation.
The semiclassical condition, $kL\gg 1$, states that the de Broglie wave
length in the relevant direction must be large in comparison to the
corresponding scale of variation of the external field $L$.
Since electroweak bubble walls are believed to be thick:
$L \sim (10-20) T^{-1}$, and relevant particle species are weakly
coupled, these constraints are satisfied for most of
particles in the electroweak plasma. Our treatment does not apply
to reflecting particles, for which $kL\sim 1$.

In a recent paper \cite{JKPl} we have analyzed the propagator of the
scalar field theory in Eq.~(\ref{Lagrange}) to non-trivial order in
gradient expansion. The main implication of this analysis is that the
quasiparticle picture, in which the plasma is treated as a collection of
one-particle excitations with a given dispersion relation, breaks down.
The reason is that in the absence of translational invariance,
states localized in coordinate space mix coherently in momentum space.
As a consequence, a self-consistent description of transport of plasma
excitations requires additional equations which encode information about
quantum coherence. In this letter we derive such a set of quantum transport
equations. Further we show that in the limit of
frequent scatterings when $\Gamma_{\! qc} L\gg 1$,
where $\Gamma_{\! qc}$ is the relevant scattering rate,
the terms describing coherent quantum effects are suppressed and can be
neglected. This can be understood simply as the result of the scattering
projecting onto the local semiclassical (quasiparticle) states. In this limit
we recover a Boltzmann equation which describes collisions and canonical
flow on non-trivially modified semiclassical trajectories in phase space.
In this letter we restrict our analysis to a simple scalar theory,
but we expect that the structure of the transport equations derived is
generic. In the discussion section we briefly outline the relevance of
our findings for computations of baryon production at the electroweak scale.

%
%
%
\section{Kadanoff-Baym Equations}
\label{sec: Kadanoff-Baym Equations}

The basic quantity in our derivation is the 2-point Green function in the
out-of-equilibrium field theory. It is most conveniently defined in the
Schwinger-Keldysh formalism \cite{SK} on a closed time contour (CTC)
\be{Gcontour}
G_{\cal C}(x,y) =   -i \left\langle T_{\cal C}
                     \left[\phi (x)\phi^\dagger (y)\right]
                 \right\rangle,
\ee
where $T_{\cal C}$ defines time ordering along the contour ${\cal C}$ which
starts at some $t_0$, often taken to be at $-\infty$, goes to $+\infty$,
and then back to $t_0$.  The two point function $G_{\cal C}(x,y)$ obeys the
contour Schwinger-Dyson equation (see Fig.~1):
\be{SDI}
G_{\cal C} (x,y) = G^0_{\cal C} (x,y)
                + \int_{\cal C} dx' \int_{\cal C} dx'' \; G^0_{\cal C} (x,x')
                  \Sigma_{\cal C} (x',x'') G_{\cal C} (x'',y)\,,
\ee
where $\Sigma_{\cal C}$ is the self-energy and $G^0_{\cal C}$ is the free
particle (tree level) propagator. In order to solve $G_{\cal C}(x,y)$ from
(\ref{SDI}) some external information about the self energy function
$\Sigma_{\cal C}$ must be provided. In general this means coupling infinitely
many new equations to (\ref{SDI}), leading to the quantum generalization
of the BBGKY hierarchy.
In the weak coupling limit it is natural to truncate the hierarchy by
substituting all higher than 2-point functions by the perturbative value
for the corresponding interaction vertex. For the scalar theory in
Eq.~(\ref{Lagrange}) the simplest such truncation consists of approximating
the four point function by the simplest four point vertex linear in the
quartic coupling $\lambda$. As a consequence the two-point self-energies are
evaluated at two loops ({\it cf.\/} Fig 1) and the resulting dynamical
equations are then truncated at the order $\lambda^2$ assuming the weak
coupling limit.
%
%
%
\begin{figure}
\leavevmode
\hspace{3truecm}
\epsfxsize=10truecm \epsfbox{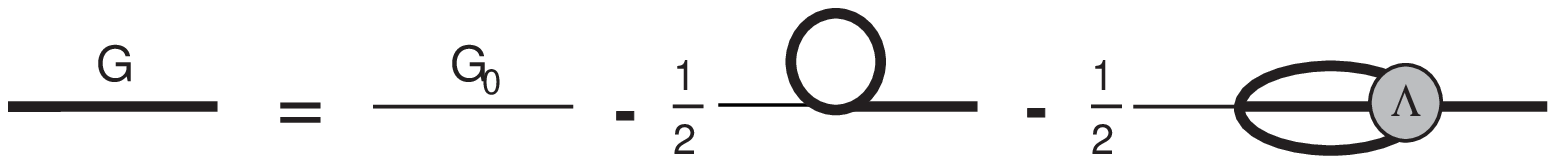}
\vspace{0.3truecm}
\baselineskip 17pt
\caption{Diagrammatic representation of the Schwinger-Dyson equation for the
             scalar theory with ${\cal L}_{\rm int}=-\lambda(\phi^*\phi)^2/4$.}
\end{figure}

The complex time ordering formulation in (\ref{Gcontour}-\ref{SDI}) can be
conveniently expressed in terms of the usual time ordered Green functions
along the real axis. If we define
\bea{GFs}
G^>(x,y)       &=& -i \langle \phi(x)\phi(y)^\dagger \rangle
       \\
G^<(x,y)       &=& -i \langle \phi(y)^\dagger\phi(x) \rangle
\label{Ggreater}    \\
G^r(x,y)       &=&  \theta(x_0-y_o) (G^>(x,y)-G^<(x,y))
\label{Gr}   \\
G^a(x,y)       &=&  -\theta(y_0-x_0) (G^>(x,y)-G^<(x,y)),
\label{Ga}
\eea
where $G^{r,a}$ are the retarded and advanced functions, we can write
(\ref{SDI}) as an equivalent set of {\em Kadanoff-Baym} equations
\cite{KB}:
\bea{KB1a}
      (G_0^{-1}-\Sigma ^{r,a})\otimes G^{r,a} (x,y) &=& \delta (x-y) \\
      (G_0^{-1}-\Sigma ^{r})  \otimes G^{<,>} (x,y) &=&
\Sigma ^{<,>}\otimes G^{a} (x,y),
\label{KB1b}
\eea
where $G_0^{-1} = \partial^2 - m^2$ is the inverse of the free propagator and
$\otimes$ represents the convolution integral:
$A\otimes B(x,y) \equiv \int d^4z A(x,z)B(z,y)$.
Functions (\ref{GFs}-\ref{Ga}) are not all independent.
In fact they can be reduced to just two independent real functions
by using (i) the hermiticity constraint $G^r(x,y)^* = G^a(y,x)$ and
$G^<(x,y)^* =-G^<(y,x)$, (ii) the relation  $G^r-G^a = G^>-G^<$ and
(iii) the spectral representation which relates $\Im G^{r,a}$ to
$\Re G^{r,a}$.

Transforming to the Wigner representation, $G(k;X) = \int d^4r e^{ik\cdot r}
G(X+r/2,X-r/2)$, and making use of the relation
$(\Sigma \otimes G) (k;X) = \exp{-i\Diamond}\DBR{\Sigma (k;X)}{G(k;X)}$,
the real part of (\ref{KB1a}) becomes an equation for the
propagator
\be{pe}
      \cos\Diamond \DBR{\Omega^2\pm i\omega\Gamma}{G^{r,a}} = 1 
\ee
and the real part of the equation (\ref{KB1b}) yields the quantum Boltzmann
equation for the dynamical variable $G^<$:
\be{qbefull}
     -\sin\Diamond\DBR{\Omega^2}{iG^<}
       \!\! = \!\! \frac{1}{2}\cos\Diamond
               \nBR{ \DBR{\Sigma^>}{G^<}- \DBR{\Sigma^<} {G^>}}
- \sin\Diamond\DBR{i\Sigma^<}{G_R},
\ee
where $\Diamond$ is the Poisson bracket operator
\be{diamond}
\Diamond\{f\}\{g\} = \frac{1}{2}\left[
                       \partial_X f \cdot \partial_k g
                     - \partial_k f \cdot \partial_X g \right]
\ee
and we defined the shorthand notation
\be{Omega}
\Omega^2=k^2-m^2-\Sigma_R.
\ee
The retarded and advanced operators were decomposed as
\be{GSigma}
G^{r,a}=G_R\mp i{\cal A}\,,\qquad
\Sigma^{r,a}=\Sigma_R\mp i\omega\Gamma,
\ee
where ${\cal A}$ is the spectral function. Finally, assuming
{\em spectral decomposition} of the Wigner functions and
making use of ${\cal A} = i(G^>-G^<)/2$, we can write
\be{spcdeco}
     iG^< = 2{\cal A} n \qquad iG^> = 2{\cal A} (n + 1).
\ee
Eqs.~(\ref{pe}--\ref{qbefull}) are the full dynamical equations in the Wigner
representation. These are suitable for the description of systems in slowly
varying backgrounds, with truncation at a given order in gradients leading
typically to a more accurate modeling of the dynamics.

\subsection{Propagator equation}

To the lowest order in gradients Eq.~(\ref{pe}) defines the familiar
propagators in a spatially constant background.  At this order the
spectral function ${\cal A}$ is singular in the limit $\Gamma \rightarrow 0$,
and defines the well known projection to the local quasiparticle on-shell.
Given the decomposition (\ref{spcdeco}), the QBE (\ref{qbefull}) also
becomes singular, allowing a reduction of the QBE
by integration over momentum (or frequency) to the
well known semiclassical Boltzmann equation involving on-shell
excitations only \cite{KB}.
When higher order gradient corrections are included, the on-shell projection
becomes more involved because of the coherent quantum effects described by
the gradient terms \cite{JKPl}. Before performing the on-shell projection for
the QBE, we shall here illustrate the technique using a simple test function.
 
Solving  Eq.~(\ref{pe}) iteratively to the lowest nontrivial order in
gradients around the lowest order pole gives
\be{G2}
       G^{r,a} \rightarrow G_2 =
        \frac{1}{z}+\frac{1}{2}\frac{m^{2\,\prime\prime}}{z^3}
       -\frac{1}{2}\frac{2k_0^2m^{2\,\prime\prime}+(m^{2\,\prime})^2}{z^4},
\ee
where $z=k_0^2-k_x^2$ with $k_0^2=\omega^2-\vec k_\parallel^{\, 2} - m^2(x)$
and we have assumed $\Sigma_R=0$ for simplicity. The spectral integral over
the momentum variable $k$ of some function ${\cal T}$ can be converted to a
contour integral encircling the multiple pole of $G$ at $z = 0$ \cite{JKPl}:
\bea{GTJ}
{\cal I}_p[{\cal T}] &\equiv&
     \frac{2}{\pi} \int_{0}^\infty dk_x {\cal A}_p {\cal T}
\nonumber\\
      &\rightarrow& {\rm Res} \left[G_p{\cal T}/\sqrt{k_0^2-z}\right]_{z=0}
      = \sum_{i=0}^{p+1} c_i {\cal T}^{(i)}(k_0),
\eea
where ${\cal T}^{(i)}(k_0)=(\partial_{k_x}^i{\cal T})(k_0)$, and
the coefficients $c_i$ may contain gradients up to $p$th order
with respect to $x$. To the leading order
in gradients ($p=0$) the spectral function
indeed becomes singular:
${\cal A}\rightarrow (\pi/2k_0)[\delta(k_x-k_0)+\delta(k_x+k_0)]$,
projecting sharply on-shell $k_x= \pm k_0$.

It should be noted that the singularity of the propagator (\ref{G2})
remains at the quasiparticle shell $k_x^2=k_0^2$. The effect of gradient
corrections in $G_p$ is to project out derivatives of ${\cal T}$ \wrt $k_x$
up to order $p+1$.  That this produces an effective shift
of the pole emerges when the contribution from the first order
derivative is included in the definition
of the projected function. To second order in gradients
\be{sceq}
{\cal I}_2[{\cal T}]  =\frac{{\cal T}(k_{\rm sc})}{k_{\rm sc}} +
c_2 {\cal T}^{(2)}(k_0) + c_3 {\cal T}^{(3)}(k_0),
\ee
where
\be{ksc}
k_{\rm sc}=k_0+\frac{1}{8}\frac{m^{2 \,\prime \prime}}{k_0^3}
+\frac{5}{32}\frac{(m^{2\,\prime})^2}{k_0^5}
\ee
is the space dependent semiclassical momentum which coincides with the
standard WKB dispersion relation.
The higher order derivative corrections in Eq.~(\ref{sceq}), however,
are of the same order in $\partial_x$ as the shift. As we will now see in
detail this leads to the breakdown of the quasi-particle approximation
when the projection is performed on the QBE.
It is then not sufficient to describe the system with a
single distribution function obtained by projecting on-shell, but some
off-shell information, represented by a finite number of derivatives of $n$,
describing coherent quantum effects, is necessary.

%
%
%
\section{Quantum Boltzmann Equation}
\label{sec: Quantum Boltzmann Equation}

We will now consider the QBE (\ref{qbefull}) to the lowest nontrivial
order
in gradients. We truncate the collision term at leading order in gradients
and focus on the structure of the flow term in the presence of a varying
background. This approximation amounts to neglecting the derivatives of the
self-energies, while retaining those of the background. (A complete
consideration that includes all second order terms in the collision
term will be given elsewhere \cite{JKP3}).  We then have
\be{qbe2}
-\left( \Diamond-\frac{1}{6}\Diamond^3 \right) \DBR{\Omega^2}{iG^<}
=\frac{1}{2}[ \Sigma^>G^< - \Sigma^<G^>].
\ee
The on-shell projection of the quantum Boltzmann equation is usually done
by
integrating over frequencies, resulting in an equation on the phase space
$\{\vec k;t,\vec x\}$ \cite{JKPt}. An alternative yet equivalent approach
of integrating over momenta is more convenient here, because in the
present
spatially varying problem the density of states is most conveniently
labeled by the conserved energy.

Let us continue with the special case $\Sigma_R = 0$. Inserting the
decomposition (\ref{spcdeco}) into (\ref{qbe2}) and writing the
$\Diamond$-terms explicitly gives
\be{Diamond}
\left( \omega \partial_t +
      \vec k \cdot \partial_{\vec x}
      +\frac{1}{2}  (\partial_x k_0^2)   \partial_{k_x}
      -\frac{1}{48}(\partial_x^3 k_0^2)\partial_{k_x}^3 \right) {\cal A} n
      =-\frac{1}{2}{\cal A} \left( i\Sigma^> n - i\Sigma^< (n+1) \right).
\ee
Integrating (\ref{Diamond}) over $k_x$ gives a dynamical equation coupling
$n(k_0)$ and its first three derivatives. The information contained
in this {\em zeroth moment} clearly does not suffice to define a closed
solution to the problem.
To provide closure we can perform integrals of Eq.~(\ref{Diamond})
weighted by some higher powers of $k_x$, in a manner analogous to the
standard derivation of fluid equations by taking moments of
the classical Boltzmann equation. There is however one crucial
difference. While for the latter case there is no control parameter
and hence no natural closure exists, in the former case the equations
close at a finite number of independent moments. This is the case because,
as shown in equation (\ref{GTJ}) above, in an integral over any
smooth test function weighted by $G_p$ only the first $p+2$ terms
are nonzero. In particular at second order in gradient expansion the closure
is obtained by the first four moments.

The QBE becomes very complicated when written in terms of $n^{(i)}(k_0)$. It
is more convenient to express the equations in terms of the weighted
projections of the generalized distribution function $n$
\be{fl}
f_l=  \theta(k_x) f^+_l +   \theta(-k_x) f^-_l,
\ee
where
\be{moments-defn}
f^+_l \equiv \frac{2}{\pi}\int_{0}^\infty dk_x k_x^l {\cal A} n \, ,
\qquad
f^-_l \equiv \frac{2}{\pi}\int_{-\infty}^0 dk_x (-k_x)^l {\cal A} n.
\ee
These functions are a straightforward generalization of the distribution
function for the spatially constant case. In the absence of gradient
corrections $f^\pm_l \rightarrow k_0^{l-1}n(\pm k_0)$, so that
the standard distribution function is then simply
$\theta(k_0) n(k_0)+\theta(-k_0) n(-k_0)$. Now using moments
(\ref{moments-defn}) it is particularly easy to prove the closure. Indeed, at
second order in gradients ${\cal A}\rightarrow {\cal A}_2$, and we have
the identity
\be{k-constraint}
\int_{0}^{\infty}dk_x
    ( k_x  -k_0)^4  k_x ^l {\cal A}_2 n = 0,
\ee
and the analogous identity holds for $k_x<0$, so that one immediately obtains
\be{k-constraint2}
f_{l+4}-4k_0f_{l+3}+6k_0^2f_{l+2}-4k_0^3f_{l+1}+k_0^4f_{l} = 0.
\ee
The constraint (\ref{k-constraint2})
allows any moment $f_l$ to be written
in terms of the four lowest ones $f_0,f_1,
f_2,f_3$. A similar binomial constraint holds at $p$-th order in gradients,
providing closure with $p+2$ moments.

It is curious to observe that the functions $f_l$ can be interpreted as
projections onto different momentum hypersurfaces ${\tilde k}_l$ of
$n$. Performing the $k$-integral in Eq.~(\ref{moments-defn})
implies that
\be{flnB}
      f^\pm_l= k_0^{l-2} \,\kappa_l \, n(\,\pm \tilde k_l)
        + {\cal O}(\partial_{k_x}^2n)_{k_x=\pm k_0},
\ee
where the $l$th shell momentum $\tilde k_l$ is given by
\be{kl}
      \tilde k_l
      = k_0+\frac{(l-1)(l-2)}{16}\,\frac{m^{2\,\prime\prime}}{k_0^3}
      +\frac{l^2-5l+5}{32}\,\frac{(m^{2\,\prime})^2}{k_0^5}
\ee
and
\be{kappa}
     \kappa_l = k_0+\frac{(l-1)(l-2)(l-3)}{48}\,
      \frac{m^{2\,\prime\prime}}{k_0^3}
      +\frac{(l-1)(l-3)(l-5)}{96}\,\frac{(m^{2\,\prime})^2}{k_0^5}.
\ee
In particular $\tilde k_0= k_0^2/\kappa_0 = k_{\rm sc}$ is the
semiclassical shell represented by the dispersion relation~(\ref{ksc}). One
thus arrives at the following intuitive picture, illustrated in Fig.\ 2:
taking moments of the QBE~(\ref{qbefull}) corresponds to discretizing momenta
with a finite set of hypersurfaces as given by Eq.~(\ref{flnB}), the flow
along which is described by the associated distribution function $f_l$.
%
%
%
\begin{figure}
\leavevmode
\hspace{4truecm}
\epsfxsize=8truecm \epsfbox{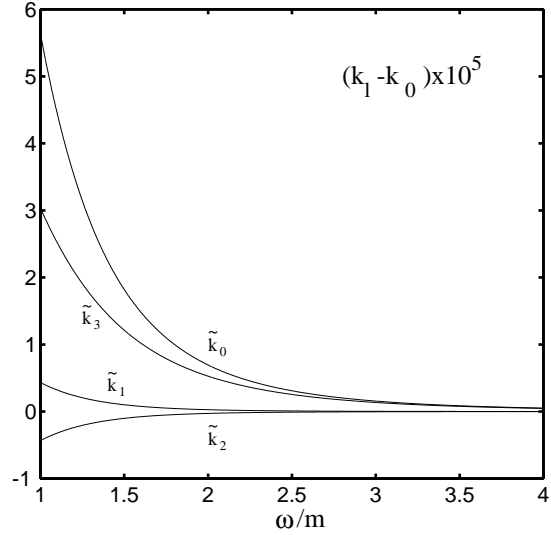}
\vspace{0.3truecm}
\baselineskip 17pt
\caption{The hypersurfaces $k_x=\tilde k_l$ on which the moments $f_l$
flow according to Eq.~(\ref{moments-be}).
}
\end{figure}

After these considerations it is easy to show that the $l\,$th moment of
QBE~(\ref{qbefull}) becomes
\be{moments-be}
\omega \partial_tf_l
      + \partial_x f_{l+1}-\frac{l}{2}(\partial_x k_0^2) f_{l-1}
      +\frac{l(l-1)(l-2)}{48}(\partial_x^3 k_0^2) f_{l-3}
       = Coll_l \,,
\ee
where we dropped the term $\vec k_\parallel\cdot \partial_{\vec
x_\parallel}f_l$
for simplicity (it can be always reinserted by the replacement
$\omega\partial_t
\rightarrow\omega\partial_t + \vec k_\parallel\cdot\partial_{\vec
x_\parallel}$).
The first four equations~(\ref{moments-be}), with $l=0,1,2,3$, together with
the closure condition~(\ref{k-constraint2}), form a
closed set for functions $f_0$, $f_1$, $f_2$ and $f_3$. These equations are our
main result. They can be used as a starting point for studying plasma dynamics
in out-of-equilibrium situations when gradient approximation applies.

Given the simple 4-point interaction term depicted in Fig.\ 1, the $l$th
moment of the collision term appearing in (\ref{moments-be}) becomes
\be{RHSpl}
     Coll_l = -\Gamma_l^{>} f_l
      +\Gamma_l^{<}\left(f_l+k_0^{l-2}\kappa_l \right)
\ee
where $\Gamma_l^>$and $\Gamma_l^<$ are the spectral projections of the
self-energies $i\Sigma^>$ and $i\Sigma^<$ defined by
\bea{Gammag-l}
\Gamma_l^{>} &=& \frac{\lambda^2}{2} \int_{p,k^\prime,p^\prime}
                    (2\pi)^{4}\delta^4 (k+p-k^\prime-p^\prime)
                    f_p (f_{p^\prime}+1) (f_{k^\prime}+1)
                    \nonumber\\
\Gamma_l^{<} &=& \frac{\lambda^2}{2}\int_{p,k^\prime,p^\prime}
                    (2\pi)^{4}\delta^4 (k+p-k^\prime-p^\prime)
                    (f_p+1)f_{p^\prime}f_{k^\prime},
\eea
where $\int_p\equiv \int d\omega d^2p_\parallel /[(2\pi)^{3}2p_{\rm sc}]$
is the semiclassical equivalent of the Lorentz invariant three dimensional
measure, and the $x$-component of the $\delta$-function should be taken as
$\delta(\theta(k_x)\tilde k_l+\theta(p_x)p_{\rm sc} -\theta(k_x^\prime)
k^\prime_{\rm sc}-\theta(p^\prime_x)p^\prime_{\rm sc})$. However, by the
same approximation we made to arrive to Eqn.\ (\ref{qbe2}), we can set
$\tilde k_l \rightarrow k_{sc}$ in the delta function, which immediately leads
to the identification $\Gamma_l^{<,>} = \Gamma_0^{<,>} \equiv \Gamma^{<,>}$.
We will see below ({\it cf.} Eq.~(\ref{change-of-var})) how, after 
a particular change of variables, $\Gamma^{<,>}$
reduces to a more canonical form. In the more general case, when second order
gradients are included, $Coll_l$ depends on moments $f_l, f_{l-1}, f_{l-2}$
and is a functional of $f_0$, but not of higher order moments, so that
including the gradient corrections to $Coll_l$ would not spoil the closure
property.

%
%
%
\section{Quantum coherence and semiclassical BE}
\label{sec: Quantum coherence and semiclassical Boltzmann equation}

In order to separate the effects of quantum coherence in
the dynamical equations (\ref{moments-be}), it is
convenient to define the following linear combinations of $f_l$:
\bea{fdelfi}
f &=& k_{\rm sc} f_0\nonumber\\
f_{\rm\! qc1} &=& f_1-f\nonumber\\
f_{\rm\! qc2} &=& f_2/\kappa_2-f_{\rm\! qc1}-f\nonumber\\
f_{\rm\! qc3} &=& f_3/k_0^2-f_{\rm\! qc2}-f_{\rm\! qc1}-f.
\eea
The distribution function $f$ measures by definition population density of
particles on the hypersurface $k_x=\tilde k_0\equiv k_{\rm sc}$, while the
densities $f_{\rm\! qci}$ measure the correlations (quantum coherence)
between neighboring hypersurfaces $k_x=\tilde k_{i-1}$ and $k_x=\tilde k_i$,
where $\tilde k_i$ are defined in Eq.~(\ref{kl}). Note that in
equilibrium in a uniform background all $f_{\rm\! qci}$ vanish.

Multiplying Eq.~(\ref{moments-be}) for $f_0$ by $k_{\rm sc}/\omega$ one
obtains
\be{f}
\partial_t f+\frac{k_{\rm sc}}{\omega}\partial_x (f+f_{\rm\! qc1})
     =-\frac{\Gamma^>}{\omega}f+\frac{\Gamma^<}{\omega}(f+1).
\ee
This equation already resembles the standard Boltzmann equation, the
main difference being coupling to an unknown function $f_{\rm\! qc1}$.
Dividing the $f_1$-equation by $\omega$ and subtracting Eq.~(\ref{f})
one then obtains
\be{f1}
(\partial_t +\Gamma_{\rm\! qc})f_{\rm\! qc1}
+\frac{\kappa_2-k_{\rm sc}}{\omega}\partial_x f_{\rm\! qc1}
+\frac{\partial_x\kappa_2}{\omega}(f_{\rm\! qc1}+f_{\rm\! qc2})
+\frac{\kappa_2}{\omega}\partial_xf_{\rm\! qc2}
=s_1,
\ee
where the source
\be{s1}
s_1 =\frac{k_{\rm sc}-\kappa_2}{\omega}\partial_x f
+\frac{(\partial_x k_0^2/2k_{\rm sc})-\partial_x\kappa_2}{\omega}f
\ee
represents coherent mixing of $f$ and $f_{\rm qc1}$. One can obtain
similar equations for the coherent quantum densities $f_{\rm\! qc2}$
and $f_{\rm\! qc3}$, but we will not present them explicitly here.
The coherent density $f_{\rm\! qci}$ is damped at the rate
$\Gamma_{\rm\! qc}$ which reads
\be{Gammaqc}
\Gamma_{\rm\! qc}=\frac{\Gamma^>-\Gamma^<}{\omega}.
\ee
This is the out-of equilibrium generalization of the on-shell damping rate
({\it cf.} Eq. ~(\ref{Gammag-l}) and \cite{Weldon}). The coherence
equations for $f_{\rm\! qci}$ are linear and hence can be quite easily solved,
and the solution for $f_{\rm\! qc1}$ inserted into Eq.~(\ref{f}).
Requiring that none of $f_{\rm\! qci}$ be sourced by the self-energy
$\Gamma^<$ defines $f_{\rm\! qci}$ uniquely.

We now pause to discuss the validity of the gradient
approximation. It is not
hard to see that in the model under study the validity criteria reduce
to the following conditions
\be{validity}
\partial_t f\ll \omega f,\qquad
\partial_x f\ll k_0 f,
\ee
which are the particular realization of $||\Diamond||\ll 1$ for the
on-shell Boltzmann equation.

In order to study how quantum coherence influences Eq.~(\ref{f}),
we now make a simple estimate of $f_{\rm\! qci}$. Not far from equilibrium
$f$ can be approximated by its equilibrium form so that
$\partial_xf\sim m^2/\omega LT$, where $L$ represents the scale on which
$m^2$ varies. To leading order in $m^2$ the source $s_1$ in Eq.~(\ref{s1})
can be estimated as
\be{qcsf2}
s_1 \sim \frac{1}{(L k_0)^3} \frac{m^2}{\omega}.
\ee
Similar estimate holds for $s_2$ and $s_3$ in the equations for
$f_{\rm\! qc2}$ and $f_{\rm\! qc3}$. If we want to include the higher order
derivatives of $k_{\rm sc}$ in Eq.~(\ref{f}), this estimate
implies that one cannot in general neglect $f_{\rm\! qc1}$ in Eq.~(\ref{f}).

There is however a limit in which it is legitimate to neglect
the part of $f_{\rm\! qc1}$ in Eq.~(\ref{f}) sourced by $s_1$
and still maintain higher order derivatives in $k_{\rm sc}$.
Indeed, assuming {\it efficient scattering}, so that coherent quantum
densities are strongly damped and the derivative terms
in Eq.~(\ref{f1}) can be neglected, we arrive at the following estimate
of the coherent quantum densities sourced by $s_1$:
\be{fqc-est}
f_{\rm\! qci} \sim
     \frac{1}{\Gamma_{\! qc} L}\frac{1}{(L k_0)^2} \frac{m^2}{\omega k_0}
\qquad (i=1,2,3).
\ee
To check consistency of this estimate, note that in the efficient scattering
limit one expects $\partial_x f_{\rm\! qci}\sim 1/L$,
so that the spatial derivative term is suppressed by $1/(\Gamma_{\rm\! qc} L)$
in comparison to the $\Gamma_{\rm\! qc} f_{\rm\! qc1}$ term. In the stationary
frame the time derivative term is suppressed by an
additional plasma velocity, which is often much smaller than unity.
For example, the phase boundary speed at the electroweak phase transition
is smaller than about 0.3 of the speed of light \cite{MP}. Upon inserting
Eq.~(\ref{fqc-est}) into Eq.~(\ref{f}), we see that the coherent quantum
density $f_{\rm\! qc1}$ is suppressed by $1/(\Gamma_{\rm\! qc} L)$ in
comparison to the other third order gradient terms in Eq.~(\ref{f}). We
have thus shown that in the limit $\Gamma_{\rm\! qc} L\gg 1$ the coherent
quantum density $f_{\rm\! qc1}$ in Eq.~(\ref{f}) can be neglected so
that we arrive at the semiclassical Boltzmann equation
\be{fC}
\partial_t f+\frac{k_{\rm sc}}{\omega}\partial_x f  = Coll[f],
\ee
where, at the leading order in gradients, the collision term is
given in Eq.~(\ref{f}). Note that this equation still does not have the
standard form since we have integrated over momentum $k_x$ and thus obtained
a distribution function $f=f(\omega,\vec k_\parallel;t,x)=k_{\rm sc}f_0$
which is a function of energy $\omega$. To recover the semiclassical
Boltzmann equation on the usual phase space we make the change of variables
defined as follows
\be{change-of-var}
\frac{d\omega}{k_{\rm sc}(\omega;x)}\rightarrow
\frac{d k_x}{\tilde\omega(k_x;x)}.
\ee
This choice is natural because it gives the local Lorentz covariant measure
in the collision integral (\ref{Gammag-l}). This change of variables in fact
does not leave $\delta^4(k+p-k^\prime-p^\prime)$ invariant. The correction
induced is, however, second order in derivatives of the distribution function
moments in the collision term, and hence it is beyond the approximation
considered in this letter. Making use of
$d\omega\rightarrow d\tilde\omega(k_x;x)=(\partial_{k_x}\tilde\omega)dk_x$,
we then immediately obtain
\be{change-of-varIII}
\partial_{k_x}\tilde\omega(k_x;x)
=\frac{k_{\rm sc}(\tilde\omega(k_x;x);x)}{\tilde\omega(k_x;x)},
\ee
which specifies $\tilde\omega=\tilde\omega(k_x;x)$.
Consider now how the flow term transforms under this change of
variables. We first have $f(\omega;x,t)\rightarrow f(k_x;x,t)$, so that
\bea{fchange}
\frac{k_{\rm sc}}{\omega}\rightarrow v_g\equiv \partial_{k_x}\tilde\omega
\nonumber\\
\partial_x\rightarrow \partial_x-(\partial_{k_x}\tilde\omega)^{-1}
(\partial_x\tilde\omega)\partial_{k_x},
\eea
and hence Eq.~(\ref{fC}) becomes
\be{fCII}
\partial_t f+v_g\partial_x f +F_x\partial_{k_x} f  = Coll[f],
\ee
with the following canonical velocity and force
\bea{vg}
v_g &=& \partial_{k_x}\tilde\omega
\nonumber\\
F_x &=& -\partial_x\tilde\omega.
\eea
This semiclassical Boltzmann equation is one of our main results. Note
that ``inversion'' of the semiclassical dispersion relation $k_{\rm sc} =
k_{\rm sc}(\omega;x)$ as given in Eq.~(\ref{change-of-varIII}) is defined
in a rather non-trivial way by invoking a natural transformation of the
collision integral measure, resulting in a different form of the
semiclassical energy hypersurface $\tilde\omega=\tilde\omega(k_x;x)$
from what one might naively guess. This is the unique choice which,
within the present approximation, renders the canonical form for the
semiclassical velocity and force~(\ref{vg}). Recall again however,
that in the present work the $\Diamond^2$-term operating on the collision
term and other analogous contributions were neglected. In a completely
consistent computation these terms must be retained, and we are currently
pursuing this computation in \cite{JKP3}. We finally emphasize that in case
when the frequent scattering  limit does not apply, one has to use more
general equations~(\ref{moments-be}) with the closure
relation~(\ref{k-constraint2}).

%
%
%
\section{Discussion}
\label{sec: Discussion}

We have generalized the semiclassical Boltzmann equation to include
the effects of a slowly varying background to nontrivial order in
gradients for a complex scalar field. Our motivation is baryon production
at a first order electroweak phase transition, where this generalization
is necessary to model CP violating effects which first appear in transport
equations beyond leading order in gradient
expansion. We found that consistent treatment of the system
to nontrivial order in gradients requires the introduction of
coherent quantum densities in addition to the usual distribution function,
and derived the corresponding dynamical equations. We also showed that
in the limit of frequent scatterings coherent quantum
effects are suppressed, and the problem reduces to a single
Boltzmann equation for the distribution function, containing a
classical force which includes gradient corrections to nontrivial order.
This is a particularly nice realization of the quantum-to-classical
transition which occurs by the dissipative dynamics within the system itself.
The dynamics is rendered dissipative by the weak coupling truncation
of the Schwinger-Dyson equation at the order $\lambda^2$, and by the
truncation of the gradient expansion at third order in gradients, resulting
in localized dynamics in the Wigner representation and irreversibility.
The rate at which the coherent quantum effects
are dissipated is simply the out-of-equilibrium damping rate~(\ref{Gammaqc}).

We now comment on how to relate our results to electroweak baryogenesis.
The relevant CP violating effects emerge generally only at non-trivial
order in gradients of the background, which is analogous to the case
considered here. To make a connection with the computation of baryogenesis
sources, it is instructive  to integrate equation~(\ref{f}) over $\omega$
and $\vec p_\parallel$ to obtain a continuity equation for the density
of particles which contains clearly separated semiclassical and quantum
mechanical sources of the form
\bea{currents}
j_{\rm sc} =  \int  \frac{d\omega d^2p_\parallel}{(2\pi)^3}
    \frac{k_{\rm sc}}{\omega}f,
\qquad
j_{\rm qc} = \int  \frac{d\omega d^2p_\parallel}{(2\pi)^3}
    \frac{k_{\rm sc}}{\omega} f_{\rm qc1}.
\eea
This should be contrasted with the situation in literature
\cite{JPT}, \cite{HN}, \cite{Riotto}, where there is
no agreement on whether the relevant CP violating sources come from
classical or quantum currents.
Moreover, in the frequent scattering limit, $\Gamma_{qc} L >>1$, we find that
the coherent quantum current $j_{\rm qc}$ is suppressed, and the
(dominant) semiclassical current then should be calculated making
use of the semiclassical transport equation~(\ref{fCII}).
This contains a classical force analogous to the one
first introduced as a source for baryogenesis
in studies of two Higgs doublet models in Ref.~\cite{JPT}, and then
subsequently applied to the MSSM in Ref.~\cite{CJK}.

Here we studied for simplicity a complex scalar
field, whereas the physically relevant cases involve mixing scalar
or fermion fields. However we believe that the novel features
we found, being essentially a consequence of localization in space, are
not specific to the complex scalar theory, but generic to all models.
How this is realized in detail in other theories is under investigation
\cite{JKPt,JKP3}.

\section* {Acknowledgements}

T.P. wishes to thank the Niels Bohr Institute and NORDITA for hospitality
and support, where a large part of his work was done. We wish to thank
D.\ B\"odeker for many useful discussions.

%
%

\nc{\ap}[3]    {{\it Ann.\ Phys.\ }{{\bf #1} {(#2)} {#3}}}
\nc{\jmp}[3]   {{\it J.\ Math.\ Phys.\ }{{\bf #1} {(#2)} {#3}}}
\nc{\np}[3]  {{\it  Nucl.\ Phys.\ }{{\bf #1} {(#2)} {#3}}}
\nc{\prep}[3]  {{\it Phys.\ Rep.\ }{{\bf #1} {(#2)} {#3}}}
\nc{\pr}[3] {{\it  Phys.\ Rev.\ }{{\bf #1} {(#2)} {#3}}}
\nc{\spjetp}[3]{{\it Sov.\ Phys.\ JETP }{{\bf #1} {(#2)} {#3}}}
\nc{\zetp}[3]  {{\it Zh.\ Eksp.\ Teor.\ Fiz.\ }{{\bf #1} {(#2)} {#3}}}

\end{document}